\documentclass[onecolumn,aps,nofootinbib,notitlepage]{revtex4-1}
\usepackage[margin=2cm]{geometry}
\usepackage{amsfonts}
\usepackage{amsmath}
\usepackage{amsthm} 
\usepackage{amssymb}
\usepackage[hidelinks]{hyperref} 
\usepackage{graphics}
\usepackage{graphicx}  
\usepackage{caption}
\usepackage{subcaption} 
\usepackage{adjustbox}
\usepackage{gensymb}
\usepackage{breqn}
\usepackage{braket}
\usepackage{gensymb} 
\usepackage[compat=1.0.0]{tikz-feynman}
 
\begin{document} 

\begin{center}
\hfill  MI-TH-1882
\end{center}

\title{\large{\textbf{ Three-loop neutrino masses via new massive gauge bosons from $E_6$ GUT}}}
\author{\normalsize{Bhaskar Dutta$^{\bf 1}$\footnote{dutta@physics.tamu.edu}, Sumit Ghosh$^{\bf 1}$\footnote{ghosh@tamu.edu}, Ilia Gogoladze$^{\bf 2}$\footnote{iliag@udel.edu}, Tianjun Li$^{\bf {3,4}}$\footnote{tli@itp.ac.cn}} \\
\vspace{1.0cm}
\normalsize\emph{$^{\bf 1}$Mitchell Institute for Fundamental Physics and Astronomy, Department of Physics  and Astronomy, Texas A$\&$M University, 
College Station, Texas 77843}\\
\normalsize\emph{$^{\bf 2}$Bartol Research Institute, Department of Physics and Astronomy,\protect \\  University of Delaware, Newark, Delaware 19716, USA}\\
\normalsize \emph{$^{\bf 3}$CAS Key Laboratory of Theoretical Physics, Institute of Theoretical Physics, \protect \\ Chinese Academy of Sciences, Beijing, 100190, People's Republic China}\\
\normalsize\emph{$^{\bf 4}$School of Physical Sciences, University of Chinese Academy of Sciences, \protect \\ Beijing 100049, People's Republic China}\\
\vspace{1.0cm} 
}
\begin{abstract} 
We propose a $SU(3)_C\times SU(2)_L \times SU(2)_N \times U(1)_Y$ model arising from  $E_6$ grand unified theory. 
We show that the tiny neutrino masses in this model can be generated at the
three-loop involving the $SU(2)_N$ gauge bosons. With Yukawa couplings around 0.01 or larger and TeV-scale $SU(2)_N$ gauge bosons, 
we show that the neutrino oscillation data can be explained naturally by presenting a concrete benchmark set of input parameters. 
All new particles are around the TeV scale. Thus  our model can be tested at the ongoing/future collider experiments.
\end{abstract}
\maketitle
\section{Introduction} 
One of the great achievements in particle physics during the last few decades 
is the discovery of the neutrino oscillations~\cite{oscillation:1,oscillation:2}, which 
can be explained by assuming nonzero masses of neutrinos.  However,  neutrinos are massless in the Standard Model (SM). Therefore,  the neutrino oscillations 
 provide a  solid evidence for new physics beyond the SM.    

The lightest charged particle in the SM is the electron, and its mass  is at least 6 orders of magnitude larger 
than the predicted neutrino mass~\cite{massrange}. Thus, any new physics theory beyond the SM should explain 
why the neutrino masses are so tiny. Several attempts have been made in last a few decades. 
In the minimal SM extension, there is a unique Weinberg's dimension five operator~\cite{Weinberg-dim-5} 
\begin{equation} 
\mathcal{L}_5= f_{ijmn}\bar{l}^C_{i \alpha L}l_{j \beta  L}\phi_\gamma^{(m)}\phi_\delta^{(n)}\epsilon_{\alpha \gamma}\epsilon_{\beta \delta}
+ f^\prime_{ijmn}\bar{l}^C_{i \alpha L}l_{j \beta  L}\phi_\gamma^{(m)}\phi_\delta^{(n)}\epsilon_{\alpha \beta}\epsilon_{\gamma \delta}~,~\,
\end{equation} 
where $\phi^{(m)}$ is scalar field and  can be one or more; $l$ is the lepton doublet; $\alpha$, $\beta$, $\gamma$ 
and $\delta$ are the $SU(2)_L$ indices; $i$ and $j$ are the generation indices. The $f$ and $f^\prime$ are roughly of 
the order $1/M$, where $M$ is the mass scale of new physics. At the tree level, there exist only three different mechanisms
to realize this operator~\cite{Ma:1}: type-I \cite{Type-I:1,Type-I:2,Type-I:3,Type-I:4}, 
type-II \cite{Type-II:1,Type-II:2,Type-II:3,Type-II:4,Type-II:5,Type-II:6}, and type-III \cite{Type-III:1} seesaw mechanisms
involving   singlet fermion, scalar triplet, and Majorana triplet fermion, respectively, as heavy intermediate particles 
with the mass of the order of $M$. We can obtain a tiny neutrino mass by integrating out the heavy fields,
which is roughly given by ${\braket{v^{(m)}_0}}^2/M$, where $\braket{v^{(m)}_0}$ is the vacuum expectation value (VEV) 
of the scalar $\phi^{(m)}$. The neutrino mass is suppressed by the heavy mass scale $M$, which is generally close to 
the unification scale in the grand unified theory (GUT) for standard high energy seesaw models where not all the Yukawa couplings are very small. Such a  high energy scale is inaccessible at  experiments like LHC.

In order to get a testable new physics scale, we need a  suppression mechanism different from the usual seesaw mechanisms. 
One such mechanism could be the radiatively generated neutrino masses~\cite{rad:1,rad:2,rad:3,rad:4,rad:5,rad:6,rad:7,rad:8,rad:9,rad:10,rad:11,rad:12,rad:13,rad:14,rad:15,rad:16,rad:17,rad:18,rad:19,loopreview}.
The suppression arises from the loop integrals and the  new physics scale $M$ is  usually the TeV scale. 
At the $n$-loop order, a dimension $d$ diagram estimates the neutrino mass as 
\begin{equation}  
m_\nu \sim c \times \left(\frac{1}{16\pi^2}\right)^n \times \frac{{\braket{v^{(m)}_0}}^{2k}}{M^{2k-1}} ~,~\,
\end{equation} 
where $c$ is a dimensionless quantity contains all the coupling constants and other mass ratios, 
and the mass dimension of the corresponding effective operator
is $2k+3$, for example, we consider the dimension-5 Weinberg operator with $n=3$ and $k=1$ in this paper.

The existing works on 3 loop masses, for example, the  KNT model \cite{rad:9}, the AKS model  \cite{rad:13} and the cocktail model \cite{cocktail}, involve new particles assuming SM gauge symmetry extended by an additional discrete symmetry. In this work, we present a model with an additional $SU(2)_N$ gauge symmetry~\cite{SU2N:1, SU2N:2} where
the gauge symmetry group $SU(2)_N$ can arise as a subgroup in the decomposition of 
the $E_6$ GUT model \cite{slansky,E6:1,E6:2,E6:3,E6:4,E6:5,E6:6}. 
The particle content of the model restricts the Majorana neutrino masses to be generated below the three-loop level. 
The $SU(2)_N$ gauge bosons play an important role in the determination of the neutrino masses  at  three loops.  
Due to a large suppression factor, $ \left(\frac{1}{16\pi^2}\right)^3$ $\sim$10$^{-7}$, arising from the loop integrals,
 the TeV mass scale can be the new physics scale of our model. The new gauge symmetry $SU(2)_N$ can be broken 
around the TeV scale, so our model can be tested at the  ongoing LHC and/or HE-LHC, FCC, and SppC, etc.

This paper is organized as follows: in Sec. II, we present the model in details  and 
 discuss the possible Yukawa coupling terms.  We study the Higgs potential and its minimization in Sec. III. 
In Sec. IV, we calculate the masses for different scalar particles and obtain the physical states. 
Sec. V includes the details about the gauge sector of the model, such as gauge boson masses and their couplings 
with scalars and fermions. We obtain an analytical expression for the neutrino mass matrix in Sec. VI. 
A numerical analysis, to show the consistency of the analytical expression with the experimental data, 
is given  in Sec. VII. LHC relevant constraints are discussed in Sec. VIII and we conclude in Sec. IX.

\section{Model Building}

Our model can arise from the $E_6$ GUT. One possible maximal subgroup of $E_6$ is $SU(6) \times SU(2)_N$. 
The $SU(6)$ group has a maximal subgroup $SU(5) \times U(1)'$. We assume that the $U(1)'$ gauge symmetry is broken
around the GUT scale. Because the $SU(5)$ group contains the SM gauge symmetry,   
the low energy gauge symmetry of our model is $SU(3)_C\times SU(2)_L \times SU(2)_N \times U(1)_Y$. 
The $SU(2)_N$  has no component to the electric charge operator in our model, so the charge operator is 
defined as $Q$ $=$ $T_{3L}+Y$. We assume that the $SU(2)_L$ doublet assignments are vertical while the  $SU(2)_N$ doublets are horizontal.

Under the gauge  symmetry $SU(3)_C\times SU(2)_L \times SU(2)_N \times U(1)_Y$,  the quantum numbers for the fermions are
{\small
 \[  Q_i\sim  \left( \begin{array}{c} u_i \\ d_i \end{array} \right) \sim (3,2,1,\frac{1}{6}) ,\ \ \ \ \  U_i^c \sim (\bar{3},1,1,-\frac{2}{3}),\ \ \ \ \  D_i^c\sim  \left( {d^\prime}_i^c \ \   d_i^c \right) \sim (\bar{3},1,2,\frac{1}{3})~,~\, \] \[  D_i\sim (3,1,1,-\frac{1}{3}),\ \ \ \ \ L_i\sim  \left( \begin{array}{cc} E_i^0 &\nu_{i} \\ E_i^- & e_i^- \end{array} \right) \sim (1,2,2,-\frac{1}{2}) , 
\ \ \ \ \  E_i^c \sim (1,1,1,1)~,~\, \]  \[  {L_i}^\prime \sim   \left( \begin{array}{c} E_i^+ \\ \bar{E_i}^{0} \end{array} \right) \sim  (1,2,1,\frac{1}{2}),\ \ \ \ \ N_i^c\sim \left( n_{1i}^c \ \ n_{2i}^c \right)\sim (1,1,2,0)~,~\,\] \[  F_i \sim  \left( \begin{array}{cc} F_{3i} & F_{1i} \\ F_{2i} & -F_{3i} \end{array} \right) \sim  (1,1,3,-1), \ \ \ \ \  {F_i}^c \sim  \left( \begin{array}{cc} {F^c_{3i}} & {F^c_{1i}} \\ {F^c_{2i}} & -{F^c_{3i}} \end{array} \right) \sim(1,1,3,1)~,~\, \] }  
where $i$  $=$ $1,2,3$.
$Q_i$, $U_i^c$, $D_i^c$, $D_i$, $L_i$, $E_i^c$, $L'_i$, and $N_i^c$ arise from the \textbf{27} representation of $E_6$,
while $F_i$ and $F_i^c$ come from the \textbf{351} and  \textbf{$\overline{\textrm{351}}$} representations of $E_6$, respectively. 
Because $E_6$ is a real group while $F_i$ and $F_i^c$ are vectorlike, the gauge anomalies are canceled.

The scalar sector of the model consists of the following particles:
{\small  \[H_d \sim \left( \begin{array}{cc} {\phi_1}^0 & {\phi_3}^0 \\ {\phi_1}^- & {\phi_3}^- \end{array} \right) \sim  (1,2,2,-\frac{1}{2}),\ \ \ \ \ H_u \sim  \left( \begin{array}{c}  {\phi_2}^+  \\ {\phi_2}^0  \end{array} \right) \sim  (1,2,1,\frac{1}{2})~,~\,  \] \[ S^0\sim \left(  S_1^0 \ \ S_2^0 \right)  \sim (1,1,2,0),\ \ \ \ \  T \sim  \left( \begin{array}{cc} {T_1}^{++} &  {T_2}^{++} \\ {T_1}^+ &  {T_2}^+ \end{array} \right) \sim  (1,2,2,\frac{3}{2})~.~\, \]  } 
The $H_d$, $H_u$, and $S^0$ come from the \textbf{27} representation, 
while the bidoublet scalar $T$ arises from the \textbf{650} representation of $E_6$.
The $SU(2)_N$ gauge symmetry is broken when $S^0$ acquires a VEV, and the electroweak gauge symmetry is broken
by the VEVs of $H_d$ and $H_u$.

The Lagrangian for the Yukawa sector and vectorlike mass terms are 
 \begin{dmath} 
-\mathcal{L}_{Yukawa} = y_{1ij}L_{i\alpha\beta}T_{\gamma \beta}F_{j\delta\delta}\epsilon_{\alpha\gamma}\epsilon_{\beta\delta}\epsilon_{\beta\delta}+   y_{2ij} L_{i\alpha\beta}{H_d}_{\gamma \beta}F^c_{j\delta\delta}\epsilon_{\alpha\gamma}\epsilon_{\beta\delta}\epsilon_{\beta\delta}+y_{3ij}Q_{i\alpha}{H_d}_{\beta\gamma}D^c_{j\delta}\epsilon_{\alpha\beta}\epsilon_{\gamma\delta}+y_{4ij}Q_{i\alpha}{H_u}_{\beta}U^c_{j}\epsilon_{\alpha\beta}+y_{5ij}D^c_{i\alpha}{S^0}_{\beta}D_j\epsilon_{\alpha\beta}+y_{6ij}{L_i}_{\alpha\gamma}{L^\prime}_{j\beta}{S^0}_{\delta}\epsilon_{\alpha\beta}\epsilon_{\gamma\delta}+y_{7ij}{L^\prime}_{i\alpha}{H_d}_{\beta\gamma}{N^c}_{j\delta}\epsilon_{\alpha\beta}\epsilon_{\gamma\delta}+y_{8ij}L_{i\alpha\gamma}{H_d}_{\beta\delta}E_j^c\epsilon_{\alpha\beta}\epsilon_{\gamma\delta}+y_{9ij}L_{i\alpha \gamma}H_{u\beta}N_{j\delta}^c \epsilon_{\alpha\beta}\epsilon_{\gamma\delta} +{ \small \frac{1}{2}M_{ij} {F}_i F^c_j} + \mu_{ij} F_i E_j^c + m_{Nij}N_i^cN_j^c~,~\,
\end{dmath}
  where $\alpha$, $\beta$, $\gamma$ and $\delta$ are $SU(2)$ indices; $i$ and $j$ are generation indices; and $\epsilon_{\alpha\beta}$ 
is the totally antisymmetric  $SU(2)$ tensor with $\epsilon_{12}$ $=$ $+1$. 
For simplicity, we assume $M_{ij}= M_i \delta_{ij}$, and $\mu_{ij} =0$. $N_i^c$s are needed (and the related terms in the above Lagrangian) 
only  if the $SU(3)_C\times SU(2)_L \times SU(2)_N \times U(1)_Y$ symmetry of our model has an  $E_6$ origin.  However, if we choose to work 
with the $E_6$ GUT model,  we introduce three pairs of vectorlike fermions $27'_i$ and $\overline{27}_i'$ under $E_6$
as well as a  discrete $Z_2$ symmetry. Under this  $Z_2$ symmetry,  $27'_i$ and $\overline{27}_i'$ are odd while
all the other particles are even. We assume the Majorana masses $m_{Nij}$ in the above equation are around the GUT scale, and then 
the type I seesaw mechanism via $N_i^c$ is highly suppressed. Moreover, we assume that the vectorlike fermions except $N_i^{c\prime}$
and  ${\overline N}_i^{c\prime}$ in $27'_i$ and $\overline{27}_i'$
have vectorlike masses around the GUT scale, and  ${\overline N}_i^{c\prime}$ have Majorana masses
around the GUT scale, while $N_i^{c\prime}$ have Majorana masses around the TeV scale. To simplify the convention, we redefine
$N_i^{c\prime}$ as $N_i^c$. Thus, we obtain the model where
only  $N_i^c$ is odd under the $Z_2$ symmetry while all the other particles are even.  In such a situation, 
the $y_{7ij}$ and  $y_{9ij}$  terms in Eq. (\ref{yukawa}) are forbidden and  the lightest fermion of $N_i^c$ can be 
a dark matter candidate. In particular, there are no low-energy neutrino mass terms at tree level.

Using the explicit components of the fields, we get 
\begin{dmath} 
\label{yukawa} -\mathcal{L}_{Yukawa} = y_{1ij}(- E_i^0 T_1^++\nu_{i}T_2^++E_i^- {T_1}^{++}-  e_i^- {T_2}^{++} )F_{3j} +   y_{2ij}(- E_i^0 \phi_1^-+\nu_{i} \phi_3^-+E_i^- \phi_1^0 -  e_i^-\phi_3^0 )F_{3j}^c+y_{3ij}[(u_i\phi_1^--d_i\phi_1^0)d^c_j-(u_i\phi_3^--d_i\phi_3^0) {d^\prime}_j^c]+y_{4ij}(u_i\phi_2^0-d_i\phi_2^+)U_j^c+y_{5ij}( {d^\prime}_i^c S_2^0-d_i^cS_1^0)D_j +y_{6ij}[(E_i^0\bar{E}_j^0-E_i^-E_j^+)S_2^0-(\nu_i\bar{E}_j^0-e_i^-E_j^+)S_1^0]+
y_{8ij}(E_i^0\phi_3^--\nu_i\phi_1^--E_i^-\phi_3^0+e_i^-\phi_1^0)E_j^c
+{ \small \frac{1}{2}M_i {F}_i F^c_i}+m_{Nij}N_i^cN_j^c~.~\,
\end{dmath} 
We consider three nonzero VEVs $\braket{\phi_1^0}$ $=$ $\frac{v_1}{\sqrt{2}}$,  
$\braket{\phi_2^0}$ $=$ $\frac{v_2}{\sqrt{2}}$, and  $\braket{S_2^0}$ $=$ $\frac{v_s}{\sqrt{2}}$.
From  Eq.~(\ref{yukawa}) we get  that  $\frac{v_1}{\sqrt{2}}$ gives the down-type quark masses 
and charged lepton masses, $\frac{v_2}{\sqrt{2}}$ gives the up-type quark masses,  and 
$\frac{v_s}{\sqrt{2}}$ gives masses to the vectorlike particle $(D_i^c, ~D_i)$, $(E_i^+,~E_i^-)$
and $(E_i^0,~\bar{E_i}^{0})$. 

The discrete $Z_2$ symmetry and the particle content of our model restricts the Majorana neutrino masses to be generated below the three-loop level. Figure \ref{fig:neutrino:int} is the three-loop diagram in the interaction basis that gives rise to the effective Majorana neutrino mass operator $L_i L_j H_d^* H_d^* /\mathcal{M}$, where $\mathcal{M}$ is some effective mass scale.


\begin{figure}[h]
\centering

\begin{tikzpicture}[scale=0.20]
\begin{feynman}
\vertex (a1) {\(L_i \)}; 
\vertex[right=1.25cm of a1] (a2);
\vertex[right=1cm of a2](a3);
\vertex[right=.75cm of a3](a4);
\vertex[right=.25cm of a4](a5); 
\vertex[right=1cm of a5] (a6);
\vertex[right=1cm of a6](a7);
\vertex[right=1cm of a7] (a8) {\(L_j\)};
\vertex[above=5em of a3](b1);
\vertex[above=5em of a6](b2);
\vertex[right=1cm of b1](b3);
\vertex[above=3em of b1](c1){\(H_d^*\)};
\vertex[above=3em of b2](c2){\(H_d^*\)};
\vertex at ($(a2)!0.5!(a3)!0.5cm!90:(a3)$) (d);

\diagram* {
(a8) -- [fermion] (a7) -- [fermion,edge label=\(L_j\)] (a6) -- [fermion,edge label=\(F_l\)] (a5)--[insertion=.04](a4)--[anti fermion,edge label=\(F^c_l\)](a3)--[anti fermion,edge label=\(L_i\)](a2)--[anti fermion](a1),
(a3)--[charged scalar,edge label=\(H_d\)](b1),
(a6)--[charged scalar,edge label=\(T\)](b2),
(b3)--[charged scalar](c1),
(b3)--[charged scalar](c2),
(a2) -- [boson, quarter left,edge label=\(X_{\mu}\)] (b1) --[charged scalar,edge label=\(H_d^*\)](b3)--[charged scalar,edge label=\(T^*\)](b2)-- [boson, quarter left,edge label=\(X_{\nu}\)] (a7),

};

\end{feynman}
\end{tikzpicture}

\caption{\label{fig:neutrino:int} Three loop diagrams in the interaction basis responsible for the Majorana neutrino masses.  }
\end{figure}

\section{The Higgs Potential}

We need the complete Higgs potential to get the physical scalar states and their masses.  
The most general renormalizable scalar potential for the Higgs scalars of our model is 
  \begin{dmath} \label{potential} 
V_{potential} =m_1^2 {H_d}^{\dag}_{\alpha\beta}{H_d}_{\beta\alpha}
+m_2^2{H_u}^{\dag}_{\alpha}{H_u}_{\alpha}
+m_s^2S^0_\alpha {S^0}^{\dag}_\alpha
+m_T^2{T}^{\dag}_{\alpha\beta}{T}_{\beta\alpha}
+\frac{\lambda_2}{2}{H_u}^{\dag}_{\alpha}{H_u}_{\alpha}{H_u}^{\dag}_{\beta}{H_u}_{\beta}
+\frac{\lambda_1}{2}{H_d}^{\dag}_{\alpha\beta}{H_d}_{\beta\alpha}{H_d}^{\dag}_{\gamma\delta}{H_d}_{\delta\gamma}
+\frac{\lambda_3}{2}{H_d}^{\dag}_{\alpha\beta}{H_d}_{\beta\gamma}{H_d}^{\dag}_{\gamma\delta}{H_d}_{\delta\alpha}
+\frac{\lambda_s}{2}S^0_\alpha {S^0}^{\dag}_\alpha S^0_\beta  {S^0}^{\dag}_\beta 
+\frac{\lambda_6}{2}{T}^{\dag}_{\alpha\beta}{T}_{\beta\alpha}{T}^{\dag}_{\gamma\delta}{T}_{\delta\gamma}
+\frac{\lambda_7}{2}{T}^{\dag}_{\alpha\beta}{T}_{\beta\gamma}{T}^{\dag}_{\gamma\delta}{T}_{\delta\alpha}
+\lambda_4{H_u}^{\dag}_{\gamma}{H_u}_{\gamma}{H_d}^{\dag}_{\alpha\beta}{H_d}_{\beta\alpha}
+\lambda_5{H_u}^{\dag}_{\alpha}{H_d}_{\alpha\beta}{H_d}^{\dag}_{\beta\gamma}{H_u}_{\gamma}
+\lambda_8{H_u}^{\dag}_{\alpha}{H_u}_{\alpha}S^0_\beta  {S^0}^{\dag}_\beta
+\lambda_9S^0_\gamma  {S^0}^{\dag}_\gamma  {H_d}^{\dag}_{\alpha\beta}{H_d}_{\beta\alpha}
+\lambda_{10} S^0_\alpha{H_d}^{\dag}_{\alpha\beta}{H_d}_{\beta\gamma} {S^0}^{\dag}_\gamma
+\lambda_{11}S^0_\gamma  {S^0}^{\dag}_\gamma{T}^{\dag}_{\alpha\beta}{T}_{\beta\alpha}
+\lambda_{12}  S^0_\alpha{T}^{\dag}_{\alpha\beta}{T}_{\beta\gamma} {S^0}^{\dag}_\gamma
+\lambda_{13} {H_u}^{\dag}_{\gamma}{H_u}_{\gamma}{T}^{\dag}_{\alpha\beta}{T}_{\beta\alpha}
+\lambda_{14} {H_u}^{\dag}_{\alpha}{T}_{\alpha\beta}{T}^{\dag}_{\beta\gamma}{H_u}_{\gamma}
+\lambda_{15} {H_d}^{\dag}_{\alpha\beta}{H_d}_{\beta\alpha}{T}^{\dag}_{\gamma\delta}{T}_{\delta\gamma}
+\lambda_{16}  {H_d}^{\dag}_{\alpha\beta}{H_d}_{\beta\gamma}{T}^{\dag}_{\gamma\delta}{T}_{\delta\alpha}
+\lambda_{17}  {H_d}^{\dag}_{\alpha\beta}{T}_{\beta\alpha}{T}^{\dag}_{\gamma\delta}{H_d}_{\delta\gamma}
+\lambda^{\prime}[{H_u}_\alpha{H_d}_{\beta\gamma}S^0_\delta \epsilon_{\alpha\beta}\epsilon_{\gamma\delta}+H.c]
+\lambda[T_{\alpha\rho}{H_d}_{\beta\sigma}{H_d}_{\gamma\mu}{H_d}_{\delta\nu}\epsilon_{\alpha\beta}\epsilon_{\rho\sigma}\epsilon_{\gamma\delta}\epsilon_{\mu\nu}+T_{\alpha\rho}{H_d}_{\beta\sigma}{H_d}_{\gamma\mu}{H_d}_{\delta\nu}\epsilon_{\alpha\beta}\epsilon_{\rho\mu}\epsilon_{\gamma\delta}\epsilon_{\sigma\nu}+H.c]~,~\, \end{dmath}  where all the parameters are real. Here $\alpha$,  $\beta$,  $\gamma$,  $\delta$,  $\rho$,  $\sigma$,  $\mu$  and  $\nu$ are the $SU(2)$ indices and $\epsilon_{\alpha\beta}$ is the totally antisymmetric  $SU(2)$ tensor with $\epsilon_{12}$ $=$ $+1$.

The minimum of the potential is given by
   \begin{dmath} 
V^0_{potential} = \frac{1}{2}m_1^2v_1^2+  \frac{1}{2}m_2^2v_2^2+ \frac{1}{2}m_s^2v_s^2+\frac{1}{8}(\lambda_1+\lambda_3)v_1^4+\frac{1}{8}\lambda_2v_2^4+\frac{1}{8}\lambda_sv_s^4+\frac{1}{4}\lambda_4v_1^2v_2^2+ \frac{1}{4}\lambda_8v_2^2v_s^2+\frac{1}{4}\lambda_9v_s^2v_1^2-\frac{1}{\sqrt{2}}\lambda^{\prime}v_1v_2v_s ~.~\,\end{dmath}  
 The minimization conditions are
   \begin{equation}m_1^2+\frac{1}{2}(\lambda_1+\lambda_3)v_1^2+\frac{1}{2}\lambda_4v_2^2+\frac{1}{2}\lambda_9v_s^2-\frac{1}{\sqrt{2}}\lambda^{\prime}\frac{v_2v_s}{v_1}=0 ~,~\, \end{equation} 
 \begin{equation}m_2^2+\frac{1}{2}\lambda_4v_1^2+\frac{1}{2}\lambda_2v_2^2+\frac{1}{2}\lambda_8v_s^2-\frac{1}{\sqrt{2}}\lambda^{\prime}\frac{v_1v_s}{v_2}=0 ~,~\, \end{equation}
 \begin{equation}m_s^2+\frac{1}{2}\lambda_9v_1^2+\frac{1}{2}\lambda_8v_2^2+\frac{1}{2}\lambda_sv_s^2-\frac{1}{\sqrt{2}}\lambda^{\prime}\frac{v_1v_2}{v_s}=0 ~.~\,\end{equation}  

After $H_d$, $H_u$ and $S^0$ acquire VEVs,  we can write them as { \small  \begin{equation} \label{hd}  H_d \sim \left( \begin{array}{cc} \frac{1}{\sqrt{2}}(v_1+\rho_1+i\eta_1) &  \frac{1}{\sqrt{2}}(\rho_3+i\eta_3)  \\ {\phi_1}^- & {\phi_3}^- \end{array} \right) ~,~\, \end{equation} 
 \begin{equation} \label{hus}  H_u \sim   \left( \begin{array}{c}  {\phi_2}^+  \\  \frac{1}{\sqrt{2}}(v_2+\rho_2+i\eta_2)  \end{array} \right) ,\ \ \  S^0 \sim \left(   \frac{1}{\sqrt{2}}(\rho_{1s}+i\eta_{1s}) \ \ \ \ \   \frac{1}{\sqrt{2}}(v_s+\rho_{2s}+i\eta_{2s}) \right) ~.~\,\end{equation} }

\section{Scalar Masses}

With the scalars in Eqs. (\ref{hd}) and (\ref{hus}), we can now obtain the terms in the Lagrangian density 
which gives masses to the different scalars from Eq.~(\ref{potential}). The mass terms for the single charged scalars are
{ \small  \begin{dmath} \label{singlecharged} 
V^{\pm}_{mass} = \left( \frac{\lambda_5v_1v_2}{2}+\frac{\lambda^{\prime}v_s}{\sqrt{2}} \right) \left( \phi_1^- \ \ \phi_2^-  \right) \left( \begin{array}{cc}\frac{v_2}{v_1} &1 \\ 1 &\frac{v_1}{v_2} \end{array} \right)  \left( \begin{array}{c} \phi_1^+ \\ \phi_2^+ \end{array} \right) 
+ 
\left( \phi_3^- \ \ T_2^-  \right) \left( \begin{array}{cc} -\frac{\lambda_3v_1^2}{2}+\frac{\lambda_5v_2^2}{2}+\frac{\lambda_{10}v_s^2}{2}+\frac{\lambda^{\prime}v_2v_s}{\sqrt{2}v_1} & 6\lambda v_1^2 \\  6\lambda v_1^2 & m_T^2+\frac{\lambda_{15}v_1^2}{2}+\frac{(\lambda_{13}+\lambda_{14})v_2^2}{2}+\frac{(\lambda_{11}+\lambda_{12})v_s^2}{2} \end{array} \right)  \left( \begin{array}{c} \phi_3^+ \\ T_2^+ \end{array} \right) 
+
\left[ m_T^2+\frac{(\lambda_{15}+\lambda_{16})v_1^2}{2}+\frac{(\lambda_{13}+\lambda_{14})v_2^2}{2}+\frac{\lambda_{11}v_s^2}{2} \right]T_1^- T_1^+  ~.~\,\end{dmath} }

  First, we get  a mixing between $\phi_1^{\pm}$  and $\phi_2^{\pm}$. That mixing gives four scalars $h_1^{\pm}$  and $h_2^{\pm}$  with mass squared zero and $\frac{v_1^2+v_2^2}{v_1 v_2} \left( \frac{\lambda_5v_1v_2}{2}+\frac{\lambda^{\prime}v_s}{\sqrt{2}} \right)$ respectively. The states are \begin{equation} \label{ps:1} h_1^{\pm}=\cos{\beta} \ \phi_1^{\pm}+\sin{\beta} \  \phi_2^{\pm}  ~,~\,\end{equation}   \begin{equation} \label{ps:2} h_2^{\pm}=-\sin{\beta} \ \phi_1^{\pm}+\cos{\beta} \  \phi_2^{\pm} ~,~\, \end{equation}
 where the mixing angle is given by $\tan{\beta}$ $=$ $\frac{v_2}{v_1}$. The two massless states  $h_1^{\pm}$ are corresponding to two charged  Goldstone modes, and the other two states  $h_2^{\pm}$ are two single charged physical scalars. 

The scalars $\phi_3^{\pm}$ and $T_2^{\pm}$ will mix and give the following four mass eigenstates:
 \begin{equation}  \label{ps:3}  H_1^{\pm}=\cos{\theta} \ \phi_3^{\pm}+\sin{\theta} \  T_2^{\pm} ~,~\, \end{equation}  
\begin{equation}  \label{ps:4}  H_2^{\pm}=-\sin{\theta} \ \phi_3^{\pm}+\cos{\theta} \  T_2^{\pm} ~,~\,\end{equation} 
with mass squared 
\begin{equation} \label{H1}  m_{H_1^{\pm}}^2=\frac{1}{2}(m^2_2+m^2_3)+\frac{1}{2}\sqrt{(m^2_2-m^2_3)^2+144\lambda^2v_1^4}  \end{equation} and  \begin{equation} \label{H2} m_{H_2^{\pm}}^2=\frac{1}{2}(m^2_2+m^2_3)-\frac{1}{2}\sqrt{(m^2_2-m^2_3)^2+144\lambda^2v_1^4}  ~,~\,
\end{equation} respectively. 
The mixing angle is given by  $\tan{2\theta}$ $=$ $\frac{12\lambda v_1^2}{m^2_2-m^2_3}$. The definition of $m^2_2$ and $m^2_3$ are
\begin{equation}\label{m2} m^2_2 = m_T^2+\frac{\lambda_{15}v_1^2}{2}+\frac{(\lambda_{13}+\lambda_{14})v_2^2}{2}+\frac{(\lambda_{11}+\lambda_{12})v_s^2}{2}~,~\, \end{equation} and \begin{equation}\label{m3} m^2_3 =  -\frac{\lambda_3v_1^2}{2}+\frac{\lambda_5v_2^2}{2}+\frac{\lambda_{10}v_s^2}{2}+\frac{\lambda^{\prime}v_2v_s}{\sqrt{2}v_1}~.~\,  \end{equation}

The four states $ H_1^{\pm}$ and $ H_2^{\pm}$ are identified as four single charged physical scalar.  
From Eq.~(\ref{singlecharged}) we get two more single charged physical scalar $T_1^{\pm}$ with mass squared 
\begin{equation}\label{T1} m_{T_1^{\pm}}^2 = m_T^2+\frac{(\lambda_{15}+\lambda_{16})v_1^2}{2}
+\frac{(\lambda_{13}+\lambda_{14})v_2^2}{2}+\frac{\lambda_{11}v_s^2}{2}~.~\,  \end{equation}

The following term of the Lagrangian density gives the masses of the double charged scalars:
{\small
\begin{dmath}  V^{\pm\pm}_{mass} =\left(T_1^{--} \ \ T_2^{--}  \right) \left( \begin{array}{cc} m_T^2+\frac{(\lambda_{15}+\lambda_{16}+\lambda_{17}) v_1^2}{2}+\frac{\lambda_{13}v_2^2}{2}+\frac{\lambda_{11} v_s^2}{2} & 0 \\ 0 & m_T^2+\frac{\lambda_{15}v_1^2}{2}+\frac{\lambda_{13}v_2^2}{2}+\frac{(\lambda_{11}+\lambda_{12})v_s^2}{2} \end{array} \right)   \left( \begin{array}{c}T_1^{++} \\ T_2^{++} \end{array} \right)~.~\,  \end{dmath} } The mass matrix is already diagonalized and gives the mass squared of the four doubly charged physical scalar $T_1^{\pm\pm}$ and  $T_2^{\pm\pm}$ \begin{equation} m_{T_1^{\pm\pm}}^2 = m_T^2+\frac{(\lambda_{15}+\lambda_{16}+\lambda_{17}) v_1^2}{2}+\frac{\lambda_{13}v_2^2}{2}+\frac{\lambda_{11} v_s^2}{2} ~,~\,\end{equation} and \begin{equation}  m_{T_2^{\pm\pm}}^2 = m_T^2+\frac{\lambda_{15}v_1^2}{2}+\frac{\lambda_{13}v_2^2}{2}+\frac{(\lambda_{11}+\lambda_{12})v_s^2}{2} ~,~\, \end{equation} respectively. Next we consider the mass terms for the five  neutral scalars
\begin{dmath}  V^{\rho}_{mass} = \left( \rho_1 \ \ \rho_2 \ \ \rho_{2s} \right)  
 \left( \begin{array}{ccc}  \frac{(\lambda_1+\lambda_3)v_1^2}{2}+\frac{\lambda^{\prime}v_2v_s}{2\sqrt{2}v_1}& \frac{\lambda_4 v_1 v_2}{2}-\frac{\lambda^{\prime}v_s}{2\sqrt{2}} & \frac{\lambda_9 v_1 v_s}{2}-\frac{\lambda^{\prime}v_2}{2\sqrt{2}} \\  \frac{\lambda_4v_1v_2}{2}-\frac{\lambda^{\prime}v_s}{2\sqrt{2}} &  \frac{\lambda_2v_2^2}{2}+\frac{\lambda^{\prime}v_1v_s}{2 \sqrt{2}v_2} & \frac{\lambda_8 v_2 v_s}{2}-\frac{\lambda^{\prime}v_1}{2\sqrt{2}} \\  \frac{\lambda_9 v_1 v_s}{2}-\frac{\lambda^{\prime}v_2}{2\sqrt{2}} &  \frac{\lambda_8 v_2 v_s}{2}-\frac{\lambda^{\prime}v_1}{2\sqrt{2}}&\frac{\lambda_sv_s^2}{2}+\frac{\lambda^{\prime}v_1v_2}{2\sqrt{2}v_s} \end{array} \right) 
  \left( \begin{array}{c}  \rho_1 \\ \rho_2 \\ \rho_{2s}  \end{array} \right) 
+
\left ( \frac{\lambda_{10}v_1v_s}{4}+\frac{\lambda^{\prime}v_2}{2\sqrt{2}}\right) \left( \rho_3 \ \  \rho_{1s}  \right) \left( \begin{array}{cc}\frac{v_s}{v_1} &1 \\ 1 &\frac{v_1}{v_s} \end{array} \right)  \left( \begin{array}{c}\rho_3 \\  \rho_{1s}  \end{array} \right) ~.~\,
 \end{dmath} Here,  $\rho_3$ and $\rho_{1s}$ are the states to mix and give one neutral scalar Goldstone mode($s_0$) and one neutral physical scalar($s_3$) with mass squared equal to  $\frac{v_1^2+v_s^2}{v_1 v_s}\left(  \frac{\lambda_{10}v_1v_s}{4}+\frac{\lambda^{\prime}v_2}{2\sqrt{2}} \right)$. We get three more neutral physical scalars($s_1$, $s_2$ and $s_{2s}$) from the mixing of $\rho_1$, $\rho_2$ and $\rho_{2s}$. The term below gives the masses of pseudoscalars
\begin{dmath}  V^{\eta}_{mass} = \frac{\lambda^{\prime}}{2\sqrt{2}}\left( \eta_1 \ \ \eta_2 \ \ \eta_{2s} \right)  
 \left( \begin{array}{ccc} \frac{v_2v_s}{v_1} & v_s & v_2  \\  v_s & \frac{v_1v_s}{v_2} & v_1\\  v_2 &  v_1 & \frac{v_1v_2}{v_s} \end{array} \right) 
  \left( \begin{array}{c} \eta_1 \\ \eta_2 \\ \eta_{2s}  \end{array} \right)
+
\left ( \frac{\lambda_{10}v_1v_s}{4}+\frac{\lambda^{\prime}v_2}{2\sqrt{2}}\right) \left( \eta_3 \ \ \eta_{1s}  \right) \left( \begin{array}{cc}\frac{v_s}{v_1} &-1 \\ -1 &\frac{v_1}{v_s} \end{array} \right)  \left( \begin{array}{c}\eta_3 \\ \eta_{1s}  \end{array} \right) ~,~\,
 \end{dmath} where $\eta_1$, $\eta_2$ and $\eta_{2s}$ will mix and give two neutral  pseudoscalars Goldstone mode($s_0^{\prime\prime}$ and $s_0^{\prime\prime\prime}$) and one physical neutral pseudoscalar($s_1^\prime$) with mass squared $\frac{\lambda^{\prime}}{2\sqrt{2}}\left( \frac{v_1^2 v_2^2+v_2^2 v_s^2+v_s^2 v_1^2}{v_1 v_2 v_s} \right)$. $\eta_3$ and $\eta_{1s}$ will mix and give another neutral pseudoscalar Goldstone mode($s_0^\prime$) and another physical neutral pseudoscalar($s_3^\prime$) with mass squared equal to $\frac{v_1^2+v_s^2}{v_1 v_s}\left(  \frac{\lambda_{10}v_1v_s}{4}+\frac{\lambda^{\prime}v_2}{2\sqrt{2}} \right)$.

We start with 24 scalar degrees of freedom and end up with 18 physical scalars. The other 6 degrees of freedom correspond to the six Goldstone mode are eaten by the massless gauge bosons. The Goldstone modes will become the longitudinal modes of gauge bosons,
which will become massive. So there will be six massive gauge bosons and one massless gauge boson. 

\section{Gauge Bosons}

In this section, we  discuss the gauge boson masses and physical gauge boson states, as well as
 their interactions with the physical scalars and fermions. The Lagrangian density,
 which gives the gauge boson masses and their interactions with the scalars, is \begin{dmath} \label{gaugescalar}  \mathcal{L}_{gauge-scalar} = \left( D_{\mu}H_u \right) ^{\dag}_{\alpha}  \left( D^{\mu}H_u \right)_{\alpha} + \left( D_{\mu}H_d^T \right) ^{\dag}_{\alpha\beta}  \left( D^{\mu}H_d^T \right)_{\beta\alpha} + \left( D_{\mu}{S^0}^T \right) ^{\dag}_{\alpha}  \left( D^{\mu}{S^0}^T \right)_{\alpha}+ \left( D_{\mu}T^T \right) ^{\dag}_{\alpha\beta}  \left( D^{\mu}T^T \right)_{\beta\alpha}~,~
 \end{dmath}  where $\alpha$ and $\beta$ are the $SU(2)$ indices. The covariant derivative is defined as 
  \begin{equation}  D_{\mu}\mathbb{I} = {\partial}_{\mu} \mathbb{I}+i\frac{g}{2}{\tau}_a W_{\mu a}+i\frac{g_2^{\prime}}{2}{\tau}_a {W^{\prime}}_{\mu a}+ig^{\prime} Y B_{\mu} \mathbb{I} ~,~ \end{equation}  
 where $g$, ${g_2^{\prime}}$ and $g^{\prime}$ are the coupling constant corresponding to the $SU(2)_L$, $SU(2)_N$, and $U(1)_Y$ groups respectively. $W_{\mu }$, $ {W^{\prime}}_{\mu }$, and $ B_{\mu}$ are the gauge bosons of the $SU(2)_L$, $SU(2)_N$, and $U(1)_Y$ groups respectively.  
\subsection{Gauge boson masses}
We define $\sqrt{2} W_{\mu}^{\pm}$ $=$ $W_{1\mu} \mp i W_{2\mu}$ and  $\sqrt{2} {X_{{1,2}\mu}}$ $=$ ${W^{\prime}}_{1\mu} \mp i{W^{\prime}}_{2\mu}$. After the spontaneous symmetry breaking of the gauge groups the massless gauge boson will become massive. We write the part of Eq.~(\ref{gaugescalar}) that  gives  the masses of the gauge bosons
  \begin{dmath}  -\mathcal{L}^{mass}_{gauge} =
 \frac{1}{4}g^2(v_1^2+v_2^2)W_{\mu}^-W^{+\mu}
+
 \frac{1}{4}{g_2^{\prime}}^2(v_1^2+v_s^2)X_{2\mu}X_1^{\mu} 
+
\frac{1}{8} \left( B_{\mu}  \ W_{3\mu} \ W^{\prime}_{3\mu} \right)  
  \left( \begin{array}{ccc}
{g^{\prime}}^2 (v_1^2+v_2^2)&-gg^{\prime} (v_1^2+v_2^2) &-g^{\prime}{g_2^{\prime}}v_1^2
\\-g g^{\prime} (v_1^2+v_2^2) &g^2(v_1^2+v_2^2)&g{g_2^{\prime}}v_1^2
 \\ -g^{\prime}{g_2^{\prime}}v_1^2&g{g_2^{\prime}}v_1^2  & {g_2^{\prime}}^2(v_1^2+v_s^2)  \end{array} \right)   \left( \begin{array}{c} B^{\mu}  \\ {W_3}^{\mu} \\ {{W}^{\prime}_{3}}^{\mu} \end{array} \right) ~.~  \end{dmath}

After the spontaneous symmetry breaking, $B_{\mu}$, $W_{3\mu}$ and $W^{\prime}_{3\mu}$ will mix and give three physical gauge bosons, which can be written as
 \begin{equation}  A_{\mu} = \sin{{\theta}_W} \ W_{3\mu} + \cos{{\theta}_W} \ B _{\mu} \end{equation}
 \begin{equation}  Z_{\mu} = \cos{{\theta}_N}  \cos{{\theta}_W} \ W_{3\mu} -\cos{{\theta}_N}  \sin{{\theta}_W} \ B _{\mu}+\sin{{\theta}_N} \ W^{\prime}_{3\mu} \end{equation}
\begin{equation} X_{3\mu} =- \sin{{\theta}_N}  \cos{{\theta}_W} \ W_{3\mu} 
+\sin{{\theta}_N}  \sin{{\theta}_W} \ B _{\mu}+\cos{{\theta}_N} \ W^{\prime}_{3\mu}  ~,~\,
\end{equation}   where the mixing angles are given by, $\tan{{\theta}_W}$ $=$ $\frac{g^{\prime}}{g}$ and $  \tan{2{\theta}_N} $ $=$ $\frac{b}{a_-}$. The definitions of $b$ and $a_{\pm}$ are 
\begin{equation} b \equiv \frac{1}{8} {g_2^{\prime}}\sqrt{g^2+{g^{\prime}}^2} v_1^2 ~,~\, \end{equation}  
\begin{equation} a_{\pm} \equiv \frac{1}{16}\left[ {g_2^{\prime}}^2(v_1^2+v_s^2) \pm (g^2+{g^{\prime}}^2)  (v_1^2+v_2^2) \right]  ~.~\,\end{equation}  
There are four other physical gauge bosons, which are $W^{\pm}_{\mu}$ and $X_{1,2\mu}$. The mass squared of all the physical gauge bosons  are then given by, 
\begin{equation} m_{W^{\pm}}^2 =   \frac{1}{4}g^2(v_1^2+v_2^2) ~,~\, \end{equation}
\begin{equation} m_{X_{1,2}}^2 =  \frac{1}{4}{g_2^{\prime}}^2(v_1^2+v_s^2) ~,~\, \end{equation}
\begin{equation} m_A^2 = 0  ~,~\, \end{equation}
\begin{equation} m_Z^2 = a_+-\sqrt{a^2_-+b^2} ~,~\, \end{equation}
\begin{equation}  m_{X_3}^2 = a_+ + \sqrt{a^2_-+b^2} ~.~\, \end{equation} 
 Also, there are exactly six massive gauge bosons corresponding to six Goldstone modes.

\subsection{Gauge bosons  interactions }

 Next, we study the interactions between the physical gauge bosons and scalars.  
A few important terms in Eq.~(\ref{gaugescalar}) are
    \begin{dmath} \label{gsint:1}  \mathcal{L}  ^{int}_{gs} =
 \frac{1}{2} {g_2^{\prime}}^2X_{2\mu}X_1^{\mu} \phi_3^+ \phi_3^-
+\frac{1}{2} {g_2^{\prime}}^2X_{2\mu}X_1^{\mu} T_2^+ T_2^-
 +\frac{i}{\sqrt{2}} {g_2^{\prime}}X_{1\mu}\left( {\partial}^{\mu}\phi_1^+ \right) \phi_3^-
 -\frac{i}{\sqrt{2}} {g_2^{\prime}}X_{1\mu} \phi_1^+\left( {\partial}^{\mu}\phi_3^- \right)
+\frac{i}{\sqrt{2}} {g_2^{\prime}}X_{2\mu}\left( {\partial}^{\mu}\phi_3^+ \right) \phi_1^- 
-\frac{i}{\sqrt{2}} {g_2^{\prime}}X_{2\mu} \phi_3^+\left( {\partial}^{\mu}\phi_1^- \right)
 +\frac{i}{\sqrt{2}} {g_2^{\prime}}X_{1\mu}\left( {\partial}^{\mu}T_1^- \right) T_2^+
 -\frac{i}{\sqrt{2}} {g_2^{\prime}}X_{1\mu}T_1^-\left( {\partial}^{\mu} T_2^+ \right)
+\frac{i}{\sqrt{2}} {g_2^{\prime}}X_{2\mu} T_1^+\left( {\partial}^{\mu}T_2^- \right)
-\frac{i}{\sqrt{2}} {g_2^{\prime}}X_{2\mu}\left( {\partial}^{\mu}T_1^+ \right)T_2^- + ...~.~\, \end{dmath} 
We  rewrite  Eq.~(\ref{gsint:1})  in terms of the physical scalars using Eqs. (\ref{ps:1})-(\ref{ps:4}),
 and then derive the necessary Feynman rules for the interactions in the following
  \begin{dmath}  \mathcal{L}  ^{int}_{gs} = 
 \frac{1}{2} {g_2^{\prime}}^2X_{2\mu}X_1^{\mu}\left[  H_1^+H_1^- +H_2^+ H_2^-   \right]
+\frac{i}{\sqrt{2}} {g_2^{\prime}}X_{2\mu}\cos{\theta} \sin{\beta}\left[  H_1^+ (\partial^\mu h_2^-)-(\partial^\mu H_1^+)h_2^-  \right]
+\frac{i}{\sqrt{2}} {g_2^{\prime}}X_{2\mu}\sin{\theta} \cos{\beta}\left[   (\partial^\mu H_2^+) h_2^-- H_2^+(\partial^\mu h_2^- ) \right]
+\frac{i}{\sqrt{2}} {g_2^{\prime}}X_{2\mu}\sin{\theta}\left[ T_1^+  (\partial^\mu H_1^-) -(\partial^\mu T_1^+)H_1^- \right]
+\frac{i}{\sqrt{2}} {g_2^{\prime}}X_{2\mu}\cos{\theta}\left[  T_1^+  (\partial^\mu H_2^-) -(\partial^\mu T_1^+)H_2^-  \right]
 +h.c +  ...~.~\,  \end{dmath} 

Next, we consider the kinetic energy terms of the $L_i$ leptons. These terms give us the interactions of the leptons with the gauge bosons. 
Let us first write down the kinetic term
\begin{dmath} \label{Lkinetic}    \mathcal{L}  ^{L}_{kinetic} =
\left(  \bar{L}_{i} \right) _{\alpha\beta} i\gamma^\mu \left(  \partial_\mu \mathbb{I} L_i \right)_{\beta\alpha}
 +\left(  \bar{L}_{i} \right) _{\alpha\beta} i\gamma^\mu \left(  \frac{1}{2}ig \tau_a W_{\mu a} L_i \right)_{\beta\alpha}
+\left(  \bar{L}^T_{i} \right) _{\alpha\beta} i\gamma^\mu \left(  \frac{1}{2}i g_2^{\prime} \tau_a W^{\prime}_{\mu a} L^T_i \right)_{\beta\alpha}
- \left(  \bar{L}_{i} \right) _{\alpha\beta} i\gamma^\mu  \left(  \frac{1}{2}i g^{\prime} B_\mu  \mathbb{I} L_i \right)_{\beta\alpha} ~,~\, \end{dmath}
 where $i$ is the generation index; $\alpha$ and $\beta$ are $SU(2)$ index; and $a$ $=$ 1,2,3. The Eq.~(\ref{Lkinetic}) 
will give us the important interaction term between the neutral leptons and gauge bosons $X_{1,2}^\mu$ as below 
\begin{equation}     \mathcal{L}  ^{L}_{kinetic} = -\frac{1}{\sqrt{2}} g_2^{\prime} X_{2\mu} \bar{\nu}_i \gamma^\mu E^0_i + H.c + ...~,~\,   \end{equation}  

We need one more interaction term which will play an important in the next section. We  write the Yukawa sector, given by Eq.~(\ref{yukawa}), 
in terms of the physical scalar and fermions. We  write all the relevant  terms here: 
\begin{equation} \mathcal{L}_{Yukawa} =- y_{1ij}E^0_i T^+_1 F_{3j} +  y_{2ij} \sin{\phi} \   E^0_i h_2^-  F_{3j}^c  + ...~.~\,\end{equation}

\section{Neutrino Masses}

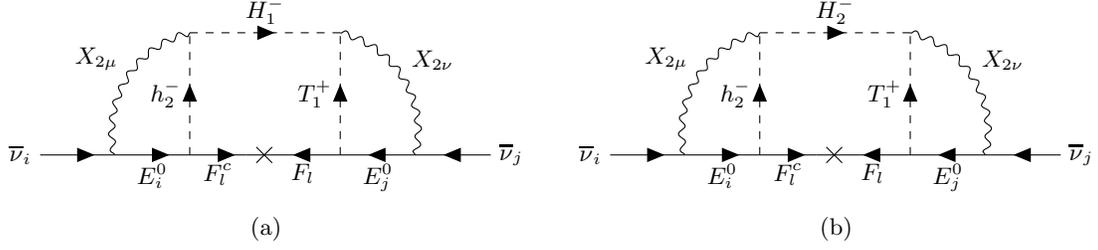
\begin{figure}[h]
\centering
  \begin{subfigure}[b]{0.4\textwidth}
        \centering
\begin{tikzpicture}[scale=0.10]
\begin{feynman}
\vertex (a1) {\(\overline \nu_i \)}; 
\vertex[right=1.25cm of a1] (a2);
\vertex[right=1cm of a2](a3);
\vertex[right=.75cm of a3](a4);
\vertex[right=.25cm of a4](a5); 
\vertex[right=1cm of a5] (a6);
\vertex[right=1cm of a6](a7);
\vertex[right=1cm of a7] (a8) {\(\overline \nu_j\)};
\vertex[above=5em of a3](b1);
\vertex[above=5em of a6](b2);
\vertex at ($(a2)!0.5!(a3)!0.5cm!90:(a3)$) (d);

\diagram* {
(a8) -- [fermion] (a7) -- [fermion,edge label=\(E^0_j\)] (a6) -- [fermion,edge label=\(F_l\)] (a5)--[insertion=.04](a4)--[anti fermion,edge label=\(F^c_l\)](a3)--[anti fermion,edge label=\(E^0_i\)](a2)--[anti fermion](a1),
(a3)--[charged scalar,edge label=\(h_2^-\)](b1),
(a6)--[charged scalar,edge label=\(T_1^+\)](b2),
(a2) -- [boson, quarter left,edge label=\(X_{2\mu}\)] (b1) --[charged scalar,edge label=\(H_{1}^-\)](b2)-- [boson, quarter left,edge label=\(X_{2\nu}\)] (a7),

};

\end{feynman}
\end{tikzpicture}
\caption{\label{fig:neutrino2}}
\end{subfigure}
\hspace{1em} 
 \begin{subfigure}[b]{0.4\textwidth}
        \centering
\begin{tikzpicture}[scale=0.10]
\begin{feynman}
\vertex (a1) {\(\overline \nu_i \)}; 
\vertex[right=1.25cm of a1] (a2);
\vertex[right=1cm of a2](a3);
\vertex[right=.75cm of a3](a4);
\vertex[right=.25cm of a4](a5); 
\vertex[right=1cm of a5] (a6);
\vertex[right=1cm of a6](a7);
\vertex[right=1cm of a7] (a8) {\(\overline \nu_j\)};
\vertex[above=5em of a3](b1);
\vertex[above=5em of a6](b2);
\vertex at ($(a2)!0.5!(a3)!0.5cm!90:(a3)$) (d);

\diagram* {
(a8) -- [fermion] (a7) -- [fermion,edge label=\(E^0_j\)] (a6) -- [fermion,edge label=\(F_l\)] (a5)--[insertion=.04](a4)--[anti fermion,edge label=\(F^c_l\)](a3)--[anti fermion,edge label=\(E^0_i\)](a2)--[anti fermion](a1),
(a3)--[charged scalar,edge label=\(h_2^-\)](b1),
(a6)--[charged scalar,edge label=\(T_1^+\)](b2),
(a2) -- [boson, quarter left,edge label=\(X_{2\mu}\)] (b1) --[charged scalar,edge label=\(H_{2}^-\)](b2)-- [boson, quarter left,edge label=\(X_{2\nu}\)] (a7),

};

\end{feynman}
\end{tikzpicture}
\caption{\label{fig:neutrino1}}
\end{subfigure}

\caption{\label{fig:neutrino} Three loop diagrams responsible for the Majorana neutrino masses. We have two similar diagrams for $X_1$ gauge boson.  }
\end{figure}

We obtain an analytical expression for the neutrino mass matrix elements  in this section. As mentioned before, 
the particle content of our model does not allow us to generate the neutrino masses below three loop;  
thus, the leading contributions to neutrino masses arise from the three loop diagrams shown in Fig. \ref{fig:neutrino}. 
The new gauge bosons, $X_1$ and $X_2$, are responsible for these three loop diagrams\footnote{We have used the package TikZ-Feynman   \cite{tikz} to draw the diagram.}.


We  have  all the necessary physical particle masses and the  interaction terms to calculate the three loop diagram in Fig. \ref{fig:neutrino}.  
In unitary gauge, the Majorana mass matrix elements are given by
\begin{dmath} \label{element}  \left( M_\nu \right)_{ji} = \frac{1}{4}{ g_2^{\prime}}^4  
y_{1jl} y _{2li} \sin{2\theta} \  {{\sin}^2{\beta}} \times  I_{3loop} ~,~\, \end{dmath} where $i$, $j$ , $l$ $=$ 1,2,3. 
And $I_{3loop}$ is the three-loop integral given by\footnote{ A large part of the loop integral calculation 
is done by using the  FeynCalc package~\cite{FeynCalc:1,FeynCalc:2}.} 
\begin{dmath}\label{loop} 
I_{3loop} = \frac{1}{(16 \pi^2)^3\left( m_X^2 - m_{0j}^2 \right)\left( m_X^2 - m_{0i}^2 \right)m_X^2} 
\int_0^\infty dr \frac{r^2}{r+M_l^2} \left[ \frac{1}{r+m_{H_1}^2} + \frac{1}{r+m_{H_2}^2} \right] \times
 \left[  4M_lm_{oj}m_{0i}\{ f_h(r,m_X^2,m_{0i}^2,m_{h_2}^2) g_{2T}(r,m_X^2,m_{0j}^2,m_{T_1}^2)+f_T(r,m_X^2,m_{0j}^2,m_{T_1}^2) g_{2h}(r,m_X^2,m_{0i}^2,m_{h_2}^2)  - m_X^2 f_h(r,m_X^2,m_{0i}^2,m_{h_2}^2) f_T(r,m_X^2,m_{0j}^2,m_{T_1}^2) \}  
+2m_{0j}f_T(r,m_X^2,m_{0j}^2,m_{T_1}^2)\{ g_{4h}(r,m_X^2,m_{0i}^2,m_{h_2}^2)-m_X^2g_{2h}(r,m_X^2,m_{0i}^2,m_{h_2}^2) \} 
-2m_{0i}f_h(r,m_X^2,m_{0i}^2,m_{h_2}^2)\{ g_{4T}(r,m_X^2,m_{0j}^2,m_{T_1}^2)-m_X^2g_{2T}(r,m_X^2,m_{0j}^2,m_{T_1}^2) \}  \right]  ~.~\, \end{dmath} 
The definitions of the  integral functions appeared in Eq.~(\ref{loop}) are
 \begin{equation}  f_h(r,m_X^2,m_{0i}^2,m_{h_2}^2) =   \int_0^1 dx \ln{\frac{x(1-x)r+(1-x)m_X^2+xm_{h_2}^2}{x(1-x)r+(1-x)m_{0i}^2+xm_{h_2}^2}} ~,~\,\end{equation}
\begin{equation}  f_T(r,m_X^2,m_{0j}^2,m_{T_1}^2) =   \int_0^1 dx \ln{\frac{x(1-x)r+(1-x)m_X^2+xm_{T_1}^2}{x(1-x)r+(1-x)m_{0j}^2+xm_{T_1}^2}}  ~,~\, \end{equation}
\begin{dmath} g_{2h}(r,m_X^2,m_{0i}^2,m_{h_2}^2) = m_X^2 \int_0^1 dx \ln {\frac{x(1-x)r+(1-x)m_X^2+xm_{h_2}^2}{m_X^2}} \protect \\ -   m_{0i}^2 \int_0^1 dx \ln {\frac{x(1-x)r+(1-x)m_{0i}^2+xm_{h_2}^2}{m_X^2}} ~,~\, \end{dmath}
\begin{dmath} g_{2T}(r,m_X^2,m_{0j}^2,m_{T_1}^2) =  m_X^2 \int_0^1 dx \ln {\frac{x(1-x)r+(1-x)m_X^2+xm_{T_1}^2}{m_X^2}} -   m_{0j}^2 \int_0^1 dx \ln {\frac{x(1-x)r+(1-x)m_{0j}^2+xm_{T_1}^2}{m_X^2}} ~,~\, \end{dmath}

\begin{dmath} g_{4h}(r,m_X^2,m_{0i}^2,m_{h_2}^2) = m_X^4 \int_0^1 dx \ln {\frac{x(1-x)r+(1-x)m_X^2+xm_{h_2}^2}{m_X^2}} \protect \\  -   m_{0i}^4 \int_0^1 dx \ln {\frac{x(1-x)r+(1-x)m_{0i}^2+xm_{h_2}^2}{m_X^2}} ~,~\, \end{dmath}

\begin{dmath} g_{4T}(r,m_X^2,m_{0j}^2,m_{T_1}^2) =  m_X^4 \int_0^1 dx \ln {\frac{x(1-x)r+(1-x)m_X^2+xm_{T_1}^2}{m_X^2}} -   m_{0j}^4 \int_0^1 dx \ln {\frac{x(1-x)r+(1-x)m_{0j}^2+xm_{T_1}^2}{m_X^2}} ~.~\, \end{dmath} 
The mass matrix elements get a large suppression from $\frac{{ g_2^{\prime}}^4}{(16 \pi^2)^3}$ $\sim$ $10^{-11}$, which pushes the new scale to TeV. The numerical analysis depends on the choice of various parameters in the model; particularly the contribution from the gauge bosons $X_{1,2}$ will be very crucial.  

 \section{Numerical Analysis} 

In this section, we show that the neutrino mass matrix given by the Eq.~(\ref{element}) can fit the neutrino oscillation data. 
We only consider the normal hierarchy of the neutrino masses. The discussion of the inverted hierarchy case 
will be similar. The best fit of the neutrino oscillation data for normal hierarchy at 3$\sigma$ range \cite{data} are  
\begin{align} \label{data}  {\sin^2{\theta_{12}}}  = & 0.271 - 0.345 ; \ \     {\sin^2{\theta_{23}}} = 0.385-0.635 ;  \ \ {\sin^2{\theta_{13}}} = 0.01934-0.02392 ; \ \  \delta_{CP} = 0^{\degree}-360^\degree ; &\notag  \\  \Delta m^2_{21} & = 7.03\times 10^{-5} \textrm{eV}- 8.09\times 10^{-5} \textrm{eV} ; \ \  \Delta m^2_{31} = 2.407\times 10^{-3} \textrm{eV}- 2.643\times 10^{-3} \textrm{eV}  ~.~\, & \end{align}
We define the  matrix $M_{d\nu}$ $=$ diag$(m_1,m_2,m_3)$ as the diagonalized neutrino mass matrix.  In the normal hierarchy scenario,  the oscillation data correspond to $m_1<m_2<m_3$. In the simplest scenario, the lightest neutrino can be assumed to be massless. We take the neutrino mass eigenvalues as follows:
\begin{equation} m_1 \simeq 0 \  \textrm{eV} ; m_2 \simeq 8.66 \times 10^{-3} \ \textrm{eV} ; m_3 \simeq  4.98 \times 10^{-2} \ \textrm{eV}  ~.~\,  \end{equation}
 We then obtain the  Majorana mass matrix from the $M_{d\nu}$ matrix as 
\begin{equation} \label{massmatrix} M_\nu = U^{-1} M_{d\nu} U ~,~\, \end{equation} 
where $U$ is the Pontecorvo-Maki-Nakagawa-Sakata (PMNS)  mixing matrix~\cite{PMNS:1,PMNS:2} parametrized by 
\begin{equation}U =  \left( \begin{array}{ccc} c_{12}c_{13} &s_{12}c_{13}&s_{13}e^{-i\delta_{CP}} \\ -s_{12}c_{23}-c_{12}s_{23}s_{13}e^{i\delta_{CP}}&c_{12}c_{23}-s_{12}s_{23}s_{13}e^{i\delta_{CP}}&s_{23}c_{13} \\ s_{12}s_{23}-c_{12}c_{23}s_{13}e^{i\delta_{CP}}&-c_{12}s_{23}-s_{12}c_{23}s_{13}e^{i\delta_{CP}}&c_{23}c_{13}  \end{array} \right) \times \mathcal{P} ~,~\, \end{equation} 
where $c_{ab}$ $ \equiv$ $\cos{\theta_{ab}}$ and  $s_{ab}$ $ \equiv$ $\sin{\theta_{ab}}$. $\mathcal{P}$ is the unit matrix for Dirac neutrinos
or a diagonal matrix with two phase angles for Majorana neutrinos. We take both the Majorana phase angles to be zero and the central values of 
the parameters from Eq.~(\ref{data}). Without loss of generality, we choose $\delta_{CP}$ to be zero as well. 
We can now obtain the mass matrix from Eq.~(\ref{massmatrix})  
\begin{equation}\label{matrix} M_\nu =  \left( \begin{array}{ccc}
6.35989\times 10^{-12}&1.18618\times 10^{-11}& 1.32647\times 10^{-11} \\1.18618\times 10^{-11} &2.3611\times 10^{-11}&2.59738\times 10^{-11} \\ 1.32647\times 10^{-11}&2.59738\times 10^{-11}&2.86893\times 10^{-11}    \end{array} \right) \ \textrm{GeV} ~.~\, \end{equation} 
 We have the constraint $v_1^2+v_2^2 \simeq 246^2$ GeV$^2$ and $\tan{\beta}=\frac{v_2}{v_1}$. The $v_1$ and $v_2$ are also constrained 
by the top and bottom quark masses. The small values of $\tan{\beta}$ will give a large value of the top quark Yukawa coupling $(y_4)_{33}$, 
which will make the model nonperturbative. To avoid that, we take $\tan{\beta}$ $=$ 2, which gives us the mixing angle $\beta$ to be equal 
to 63\degree. We then  get $\frac{v_1}{\sqrt{2}}$ and $\frac{v_2}{\sqrt{2}}$  to  be 78  GeV and  156 GeV, respectively. 
We choose the VEV  $\frac{v_s}{\sqrt{2}}$ at the TeV scale,  to  be 17 TeV. 
Now choosing $g_2^\prime$ to be equal to 0.35, we get the mass of the gauge boson $X$ to be 5 TeV. By choosing appropriate values for different $\lambda $ parameters in Eqs.~(\ref{m2}) and (\ref{m3}), we can take $m^2_2$ and $m^2_3$ to be  2.5 $\times$ $10^7$  GeV$^2$ and 2.5 $\times$ $10^5$  GeV$^2$ respectively. Now choosing $\lambda$ $\simeq$ .03, we obtain  $m_{H_1}$ and $m_{H_2 }$ to be 5 TeV and 500 GeV respectively  from Eqs.~(\ref{H1}) and  (\ref{H2})  using the mixing angle, $\theta$$=$ 0.005\degree. Similarly,  Eq.~(\ref{T1})  gives  $m_{T_1}$ to be 500 GeV.  The other mass parameters needed are the vectorlike particle masses. The lower bound on the  vectorlike lepton mass is 101 GeV, which comes from the LEP experiment~\cite{pdg}. We take $m_0$ to be 115 GeV, 125 GeV,  and 135 GeV respectively for the first, second, and third generations. We use  M, the mass for the $F$ particle to be 110 GeV, 120 GeV, and 130 Gev respectively for the three generations. All  parameters are as follows:
\begin{align} \label{parameter} m_{H_1} =  & 5   \ \ \textrm{TeV}, \ \    m_{H_2} =   500 \ \ \textrm{GeV},   \ \ m_X = 5 \ \ \textrm{TeV}, \ \  m_{h_2} = 268 \ \ \textrm{GeV},  \ \ m_{T_1} = 500 \ \ \textrm{GeV} ~,~\,  &\notag  \\  \beta & =63\degree, \ \ \theta = 0.005\degree, \ \ M = (110,120,130) \ \ \textrm{GeV}, \ \ m_0 = (115,125,135) \ \ \textrm{GeV} ~.~\, &  \end{align}
We can use these  parameters in  Eq.~(\ref{element}) to fit the neutrino mass matrix given in Eq.~(\ref{matrix}) for Yukawa couplings that satisfy  $y_1$ $\times$  $y_2$ to be of the order of 0.001 to 0.0001. 

We have presented one set of viable input parameters. There exist many  other possible sets as well. 
The Yukawa coupling constants  can be made even larger by taking larger 
values of  $\frac{v_s}{\sqrt{2}}$. The mass $ m_X$ controls the value of the numerical integration. 
The other mass parameters do not play an important role in the calculations. The  mass gap between $m_{H_1}$ and $m_{H_2}$ can be small,
which does not affect the numerical result in any significant way.  
Another important factor, which affects the value of the numerical result, is the loop suppression factor. 
The value of $\tan{\beta}$ can  change the numerical results as well.

\section{Relevant LHC constraints }

In the previous section, we have shown, by presenting a set of parameters, that the analytically obtained neutrino mass matrix can fit the experimental data on neutrino oscillation. In this section, we discuss the constraints on the masses of the particles, that appear in the three-loop diagram that generates the neutrino mass, from the LHC searches. 

In the following table we write the different possible final states of the particles that can be observed in the LHC or any future collider experiments.


\begin{table}[h]
\centering
\begin{tabular}{l | l }
Particles  \ \ &  \ \ Possible final states \\
 \ \  &  \ \  \\
\hline

$X_{3}^{\mu}$ & \ \  $e_i^+e_i^- $ \\ 
&  \ \ $d_i\bar{d_i}$ \\

\hline

$H_1^- / H_2^-$ &  \ \  $u_i\bar{d_j}+{\nu_i}+{E^0_j}$ \\
 & \ \  $d_i\bar{d_j}+{e_j^-}+{E^0_i}$ \\
 
 \hline
 $h_2^-$& \ \ $ u_i\bar{d_j}+{E^0_i}+{E^0_j}$ \\ 
 & \ \ $ u_i\bar{d_j}$\\
 
 \hline
 $T_1^+$ & \ \ $\bar{u_i}d_j+E^0_i+\bar{E^0_j}$ \\
 & \ \ $ d_i\bar{d_j}+{e_i^+}+{E^0_i}+\bar{\nu_i}+\bar{E^0_i}$ \\
 \hline
 $F_3^c$ & \ \ $u_i\bar{d_j}+E_i^0$ \\
 & \ \ $ d_i\bar{d_j}+{e_i^+}+\bar{E^0_i}+\bar\nu_i$ \\
 \hline
\end{tabular}
\caption{ The list of the fields needed for calculating the three-loop diagram and their signatures.
}
\label{table:1}
\end{table}


The gauge bosons $X_1^\mu$, $X_2^\mu$ and $X_3^\mu$ associated with the new gauge group $SU(2)_N$ have similar masses. $X_1^\mu$ and $X_2^\mu$ gauge bosons can not be produced directly in the LHC, but the other gauge boson $X_3^\mu$(which is like $Z^\prime$) can be produced as $d_i\bar{d_i} \rightarrow X_3^\mu$. The experimental bounds on the gauge boson masses can be found by looking into the decay channel of $X_3^\mu$, which is $X_3^\mu \rightarrow e_i^+e_i^-$. Heavy neutral particle decaying into the dilepton final state has been searched by both the ATLAS~\cite{Dilepton:ATLAS} and CMS~\cite{Dilepton:CMS2016,Dilepton:CMS2017} Collaborations using 36 fb$^{-1}$ of proton-proton collision data at $\sqrt{s}=13$ TeV at the LHC.  We choose $g_2^\prime$ to be $\sim$0.35. 
Using this coupling, we find the lower limit of $X_3^\mu$ mass to be  around 3.8 TeV. The other decay channel has the dijet final state. The search for a dijet resonance final state by CMS~\cite{CMS:Dijet} and ATLAS~\cite{ATLAS:Dijet} in the proton-proton collision at $\sqrt{s}=13$ TeV corresponding to an integrated luminosity up to
36 fb$^{-1}$ and 37 fb$^{-1}$ respectively, gives a lower limit of the $X_3^\mu$ mass to be around 3.6 TeV for the coupling constant $g_2^\prime$ to be $\sim$0.35. We have used $m_X$=5 TeV for our three-loop calculation.

The scalar ($H$, $h_2$, $T_1$)-quark interactions arise from the Yukawa couplings which are related to the up-type and down-type qaurk mass generations, and therefore the couplings have hierarchy. The third generation couplings dominate. Similarly the Yukawa coupling constant for the third generation lepton is dominant. Therefore, the scalars are produced by Drell-Yan process at the LHC. Further, these new particles that appear in the loop calculation mostly decay into the lightest neutral particle $E^0_i$; the final states contain missing energy which means that the current constraints arising from the supersymmetry (SUSY) particle searches in R-parity conserving models can be applied to these new particles. 

The heavy charged scalars $H_{1,2}^{\pm}$, $h_2^\pm$ and $T_1^\pm$ are produced in pairs via the Drell-Yan process  at the LHC. We can obtain the constraints on the charged scalar mass from the final states of the charged scalar pair decay products as shown in Table \ref{table:1}. The final states of $H_{1,2}^{\pm}$ can be obtained from $pp\rightarrow H_1^-H_1^+/H_2^-H_2^+(H_1^-/H_2^-\rightarrow \nu_i F^c_{3j}, F^c_{3j}\rightarrow E_j^0h_2^-, h_2^- \rightarrow u_i\bar{d_j})$ and $pp\rightarrow H_1^-H_1^+/H_2^-H_2^+(H_1^-/H_2^-\rightarrow E_i^0E_j^c, E^c_{j}\rightarrow e^-_j s_2, s_2\rightarrow d_i\bar{d_j})$. For $h_2^\pm$, $pp\rightarrow h_2^-h_2^+(h_2^-\rightarrow E^0_i F^c_{3j}, F^c_{3j}\rightarrow E_j^0h_2^-, h_2^- \rightarrow u_i\bar{d_j})$ and $pp\rightarrow h_2^-h_2^+(h_2^-\rightarrow u_i\bar{d_j})$ give the final states. We get the final states of $T_1^\pm$ from $pp \rightarrow T_1^+T_1^-(T_1^+\rightarrow E_i^0F_{3j}, F_{3j}\rightarrow\bar{E_j^0}h_2^+, h_2^+\rightarrow\bar{u_i}d_j)$ and $pp \rightarrow T_1^+T_1^-(T_1^+\rightarrow E_i^0F_{3j}, F_{3j}\rightarrow\bar{\nu_i}H_1^+, H_1^+\rightarrow\bar{E^0_i}E^c_j, E^c_j\rightarrow e_i^+s_2, s_2\rightarrow d_i\bar{d_j} )$. 

The final states can be multijet plus missing transverse energy (MET) or multijet plus multi $e,\,\mu,\,\tau$ plus MET where MET can be used to suppress the SM background. The multijet plus MET final state is similar to the final states arising from the  gluino pair production   at LHC. The data collected by CMS~\cite{CMS:gluino}  at $\sqrt{s}=13$ TeV, with an integrated luminosity of 35.9 fb$^{-1}$ and by ATLAS~\cite{ATLAS:gluino} at $\sqrt{s}=13$ TeV, with a luminosity of 36 fb$^{-1}$ constrain multijet plus MET production cross section to be $\leq$ 1 fb. 
The charged scalar crosssection is around 0.1 fb $\times$ branching ratio  for a 500 GeV charged scalar mass ($H_{1,2}^{\pm}$ and $T_1^\pm$) used in our loop diagram~\cite{ChargedScalar},  which is  well below the current limit. 

We also have  multijet plus oppositely charged  taus plus MET for $H_{1,2}^{\pm}$, $h_2^\pm$ and $T_1^\pm$  and multijet plus oppositely charged leptons ($e,\,\mu$) plus MET  for $H_{1,2}^{\pm}$,  and $T_1^\pm$, where multileptons final states arise from the top quark decay. Data collected with the CMS~\cite{CMS:MJMLMET} at $\sqrt{s}=13$ TeV and integrated luminosity of 35.9 fb$^{-1}$ and by ATLAS~\cite{ATLAS:MJMLMET} at $\sqrt{s}=13$ TeV and integrated luminosity of 36.1 fb$^{-1}$  search for new physics with a  final state containing two oppositely charged, same flavored leptons($e^+e^-$ and $\mu^+\mu^-$), multijets and MET~\cite{CMS:MJMLMET} and 
 the production cross section is constrained to be  $\leq$ 1 fb, however, as mentioned before, the charged scalar ($H_{1,2}^{\pm}$ and $T_1^\pm$) pair production cross section is 0.1 fb for 500 GeV mass as used in the loop calculation.  The final states of $H_{1,2}^{\pm}$, $h_2^\pm$ and $T_1^\pm$ can contain multijet plus oppositely charged  taus plus MET. Final states with $tau$ are less constrained. The search for chargino  pair production with the final state involving multitau($\tau^+\tau^-$) by CMS~\cite{CMS:DiTau}(with an integrated luminosity 35.9 fb$^{-1}$)  and ATLAS~\cite{ATLAS:DiTau} (with an integrated luminosity 36.1 fb$^{-1}$) in the proton-proton collision at $\sqrt{s}=13$ constrain the production cross section to be $\leq$ 10 fb. However, the production crosssections for 300 GeV $h_2^\pm$ and 500 GeV $H_{1,2}^{\pm}$ and $T_1^\pm$ (as used in the calculation) are 0.3 and 0.1 fb respectively   Therefore, we do not get any constraint on the charged scalar masses used in the loop calculation from the final states involving multitau and multilepton ($e,\,\mu$).
 
The vectorlike lepton $F_3^c$ is produced in pair in Drell-Yan process in proton-proton collision at the LHC. The $F_3^c$ pair decays to the lightest neutral particle $E_i^0$ and jet. $pp\rightarrow F^c_3i F_{3i}(F^c_3i\rightarrow E_i^0h_2^-, h_2^- \rightarrow u_i\bar{d_j})$ and $pp \rightarrow F^c_3i F_{3i}(F_3j\rightarrow\bar{\nu_i}H_1^+, H_1^+\rightarrow\bar{E^0_i}E^c_j, E^c_j\rightarrow e_i^+s_2, s_2\rightarrow d_i\bar{d_j} ) $ give the final states. The typical masses for $F_3^c$ and $E_i^0$ are chosen to be 110-130 GeV and 115-135 GeV  for the loop calculation. Since the mass of $E_i^0$ is close to the mass of $F_3^c$, the direct production of $F_3^c$ does not contain enough missing energy  along with  soft taus and jets in the final state which makes it extremely difficult to extract the signal from the SM background. This problem is similar to the search for chargino-second lightest neutralino with smaller mass gaps between the charginos/second lightest neutralino and the lightest neutralino. In fact, the final states of the vectorlike leptons  are similar   to the chargino-second lightest neutralino pair production where the chargino/second lightest neutralino decay to the lightest neutralino with W/Z bosons in the final states. Search for chargino-second lightest neutralino production at ATLAS~\cite{ATLAS:VECLEPTON} and CMS~\cite{CMS:VECLEPTON} at $\sqrt{s}=13$ TeV and integrated luminosity of 36.1 fb$^{-1}$ and 35.9 fb$^{-1}$ respectively, show that there is no constraint for mass difference $\leq$ 70 GeV for any chargino/second-lightest neutralino mass. Further, these experimental searches used final states with $e,\,\mu$. However, if the final state dominantly contains taus (our case) we do not have any constraint from the LHC. It is possible to have jets plus missing energy final state from the $F_3$ final state (\ref{table:1}), but jets are too soft due to the small mass gap, and there is not enough missing energy. Therefore, we do not have any constraint on the vectorlike lepton masses which we have used in our loop calculation from the LHC as of yet.

\section{Conclusion} 

To construct a natural radiative neutrino mass model which can be tested at the future collider experiments,
 we have extended the SM gauge symmetry by the $SU(2)_N$ gauge group, which comes from the decomposition of the $E_6$ GUT. 
We have  presented the  particle content and all the possible Yukawa interactions and
 studied the scalar and gauge sectors in details.  Interestingly, 
the tiny neutrino masses are found to be only  generated at 
three loops where the $SU(2)_N$ gauge bosons play an important role.
 The new gauge bosons $X_{1,2}$  and vectorlike fermions enter into  the three loop diagrams. 
Because of the large suppression from the loop integral, the new physics scale can be around TeV, which is testable, 
unlike the high-scale tree-level seesaw mechanism, as well as the one-loop and two-loop neutrino mass models.

We have obtained an analytical expression for the Majorana neutrino masses. This mass expression depends on the spontaneous symmetry breaking 
scale of the $SU(2)_N$ gauge group. From the three loop calculation,  we  have shown that the analytical expression, in our radiative neutrino mass model,  
is consistent with the neutrino oscillation data. For example, for $\frac{v_s}{\sqrt{2}}$ to be 17 TeV and the new gauge boson mass to be 5 TeV, the other mass parameters are chosen to be between the electroweak  and TeV scale, which is consistent with our goal of obtaining neutrino mass at experimentally testable scale. Using these input parameters along with  the neutrino mass matrix  obtained from the oscillation data, we found
the Yukawa couplings to be 0.01 or larger. 
For larger values of $\frac{v_s}{\sqrt{2}}$ the Yukawa couplings will be larger. The typical masses of the new particles used in our three-loop calculation are allowed by the current LHC data. 

 \textbf{Acknowledgments}

BD and SG are supported in part by the DOE Grant No. DE-SC0010813. IG is supported in part by Bartol Research Institute. TL is supported in part by the Projects No. 11475238 and No. 11647601 supported by National Natural Science Foundation of China and by Key Research Program of Frontier Science, CAS.

\bibliographystyle{plain}

\end{document}